%% file: N157B_AandA.tex
\documentclass[traditabstract,longauth,letter]{aa}

\usepackage{graphicx}
\usepackage{txfonts}
\usepackage{natbib}
\bibpunct{(}{)}{;}{a}{}{,} 

\newcommand{\nb}{N\,157B}
\newcommand{\psr}{PSR\,J0537$-$6910}
\newcommand{\hess}{HESS\,J0537$-$691}
\newcommand{\hessTel}{H.E.S.S.}

\newcommand{\PhiZero}{$(8.2\pm 0.8_\mathrm{stat}) \times 10^{-13}\,\mathrm{cm}^{-2}\mathrm{s}^{-1}\mathrm{TeV}^{-1}$}
\newcommand{\PhiZeroSys}{$(8.2\pm 0.8_\mathrm{stat}\pm 2.5_\mathrm{syst}) \times 10^{-13}\,\mathrm{cm}^{-2}\mathrm{s}^{-1} \mathrm{TeV}^{-1}$}
\newcommand{\SpecInd}{$2.8 \pm 0.2_\mathrm{stat}$}
\newcommand{\SpecIndSys}{$2.8 \pm 0.2_\mathrm{stat} \pm 0.3_\mathrm{syst}$}
\newcommand{\EFlux}{$(1.4 \pm 0.1)\times 10^{-12} \, \mathrm{erg \, cm}^{-2} \, \mathrm{s}^{-1}$}
\newcommand{\CrabUnits}{2\%}
\newcommand{\Luminosity}{$(3.9 \pm 0.3) \times 10^{35} d^2_{48} \, \mathrm{erg \, s}^{-1}$}
\newcommand{\efficiency}{$0.08\% \pm 0.01\%$}
\newcommand{\efficiencyDist}{($0.08 \pm 0.01)d^2_{48}\%$}

\newcommand{\Non}{395}
\newcommand{\Noff}{3152}
\newcommand{\BGalpha}{18.62}
\newcommand{\Nex}{226}
\newcommand{\Nsig}{$14\,\sigma$}

\defcitealias{GCposition}{Acero et al. 2010}
\defcitealias{HESS_crab}{Aharonian et al. 2006a}
\defcitealias{GPS}{Aharonian et al. 2005}
\defcitealias{GPS2}{Aharonian et al. 2006b}
\defcitealias{GPS3}{Gast et al. 2012}

\input{includes}

\begin{document}
%

\title{Discovery of gamma-ray emission from the extragalactic pulsar wind nebula \nb\ with the High Energy Stereoscopic System}

\authorrunning{H.E.S.S. Collaboration}
\titlerunning{Gamma-ray emission from the extragalactic pulsar wind nebula \nb}

\input{authorsAandA}

\offprints{Nukri Komin (komin@lapp.in2p3.fr)}

   \date{Received 27/06/2012; accepted 06/08/2012}

 
  \abstract
{We present the significant detection of the first extragalactic pulsar wind nebula (PWN) detected in gamma rays, \nb, located in the large Magellanic Cloud (LMC).
Pulsars with high spin-down luminosity are found to power energised nebulae that emit gamma rays up to energies of several tens of TeV. \nb\ is associated with \psr, which is the pulsar with the highest known spin-down luminosity. 
The High Energy Stereoscopic System telescope array observed this nebula on a yearly basis from 2004 to 2009 with a dead-time corrected exposure of 46\,h. 
The gamma-ray spectrum between 600\,GeV and 12\,TeV is well-described by a pure power-law with a photon index of \SpecIndSys\ and a normalisation at
1\,TeV of \PhiZeroSys. A leptonic multi-wavelength model shows that an energy of about $4\times10^{49}\mathrm{erg}$ is stored in electrons and positrons. 
The apparent efficiency, which is the ratio of the TeV gamma-ray luminosity to the pulsar's spin-down luminosity, \efficiency, is comparable to those of PWNe found in the Milky Way. 
The detection of a PWN at such a large distance is possible due to the pulsar's favourable spin-down
luminosity and a bright infrared photon-field serving as an inverse-Compton-scattering target for
accelerated leptons.
By applying a calorimetric technique to these observations, the pulsar's birth period is estimated to be shorter than 10 ms.
}

   \maketitle
%

\section{Introduction}

In recent years, many Galactic pulsar wind nebulae (PWNe) have been discovered to be
gamma-ray emitters \citep[for a review, see e.g.][]{Emma}. 
It was predicted by \citet{Aharonian1997} that the gamma-ray
luminosity of these nebulae is connected to the spin-down
power $\dot{E}$, i.e. the loss rate of rotational energy of the pulsar, and that
pulsars with $\dot{E}
> 10^{34 - 35} (d/\mathrm{1 kpc})^2 \, \mathrm{erg\,s}^{-1}$ (with $d$ being the distance to the pulsar)  power nebulae that are detectable in
gamma rays. On the basis of the Galactic Plane Survey carried out by \hessTel\ 
\citepalias{GPS,GPS2,GPS3}
, \citet{PSRstat} suggested that pulsars with a spin-down
power of $\dot{E}
\gtrsim 10^{34} (d/\mathrm{1 kpc})^2 \, \mathrm{erg\,s}^{-1}$ may be correlated with nebulae detectable by \hessTel

The most energetic pulsar known is \psr, with a spin-down power of $\dot{E} = 4.9 \times 10^{38} \, \mathrm{erg \,
  s}^{-1}$ \citep{Marshall1998,ATNF}. \psr\ is also one the most distant pulsars known to date: located in the LMC, it has an estimated distance of $(48.1 \pm 2.3_\mathrm{stat} \pm 2.9_\mathrm{syst})$\,kpc \citep{Macri2006}. This pulsar is the compact central object of the
supernova remnant (SNR) \nb\ (also called LHA 120-N 157B, 
SNR\,B0538$-$691, or NGC\,2060). \nb\ has been observed extensively in X-rays \citep{WangGotthelf1998,Wang2001,Chen}.
\nb\ is a Crab-like SNR \citep{1996rftu.proc..255G}: the emission from \nb\ is dominated by synchrotron radiation from the PWN, which shows a bar-like feature surrounding the pulsar representing the reverse shock of a toroidal wind from the pulsar as well as a long tail of diffuse emission of about $20\arcsec \times 30\arcsec$ in the north-west direction from the bar \citep{Wang2001,Chen} (see also
Fig.~\ref{Fig1}b). 
\citet{Chen} show that faint thermal emission with a diameter of about $100\arcsec$ is observed from the supernova ejecta and, unlike shell-type or composite SNRs, no emission from the supernova forward shock is detected.
The supernova ejecta apparently expand into the low-density interior of the superbubble formed by the stellar association LH\,99 \citep{Chen}.
The characteristic age of the pulsar is about 5000 years \citep{Marshall1998}, which is consistent with estimates of the SNR's age \citep{WangGotthelf1998}. Large-scale diffuse emission from the general direction of \nb\ was discovered with the \textit{Fermi} satellite \citep{FermiLMC}, which was interpreted as emission from massive star-forming regions. No significant emission, in addition to the diffuse emission, has been detected from \nb.

Despite its extreme distance, \psr\ has a large enough spin-down flux of $\dot{E}/{d^2} = 2 \times 10^{35}\, \mathrm{erg\,s}^{-1}\mathrm{kpc}^{-2}$ to power a nebula detectable with \hessTel

\section{High Energy Stereoscopic System observations and results}

\begin{figure*}
   \centering
   \includegraphics[width=\textwidth]{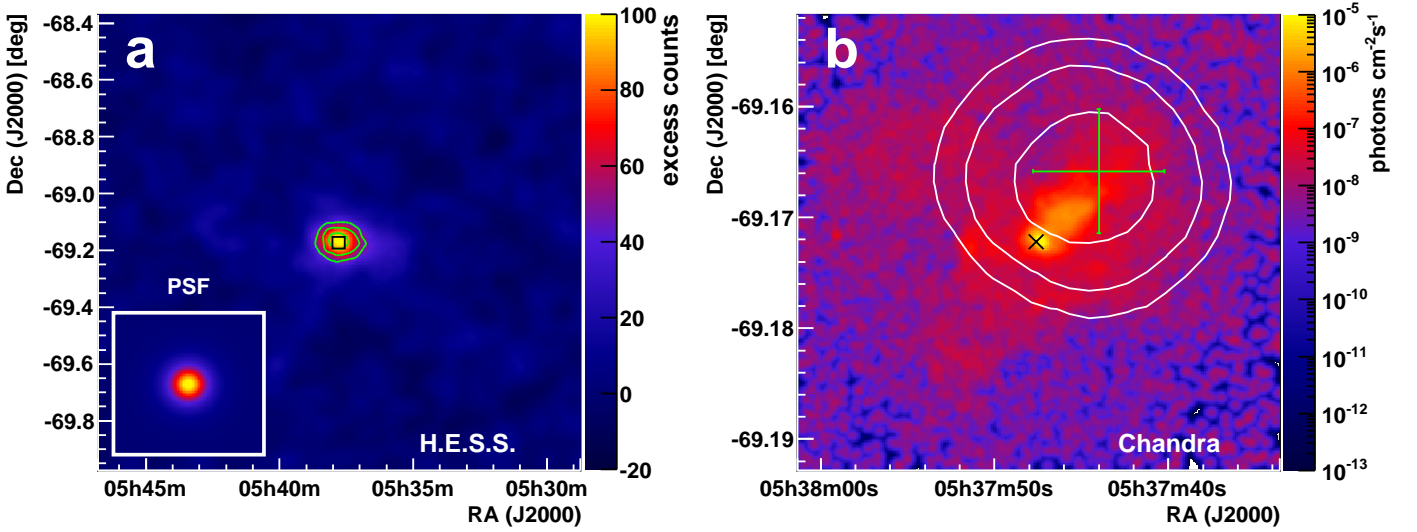}
   \caption{Gamma-ray and X-ray view of \nb.
	{\bf a)} Image of the TeV
  gamma-ray excess obtained with H.E.S.S. The image was smoothed with
the instrument's point-spread function, which is shown in the inset.
  The
  green contour lines denote 6, 9, and 12 $\sigma$ statistical
  significances. The black box outlines the region shown in the X-ray image.
{\bf b)} 
Image of the X-ray flux in the energy band from 0.8\,keV to 8\,keV (\textit{Chandra} ACIS-S, observation n$^{\circ}$2783, 48.2 ks dead-time corrected exposure). The data were smoothed with a Gaussian of $\sigma=1.5\arcsec$ and the exposure map was computed with an 
assumed spectral index of 2.3, which is the average index of the nebula.
  The colour coding represents the X-ray flux
  in photons cm$^{-2}$ s $^{-1}$; for better visibility, it is clipped
  at $10^{-13}\,\mathrm{photons\,cm}^{-2}\, \mathrm{s}^{-1}$ at the lower end
  and saturates at $10^{-5}\,\mathrm{photons\,cm}^{-2}\, \mathrm{s}^{-1}$.  The
  white contour lines denote regions of 68\%, 95\%, and 99\% confidence for
  the position of the gamma-ray source {\hess}. 
  The position of the pulsar \psr\
  is indicated by a black cross. The green cross marks the best-fit position and the systematic uncertainty in the pointing of the H.E.S.S. telescopes.
   }
              \label{Fig1}%
    \end{figure*}

The High Energy Stereoscopic System (\hessTel) \citepalias{HESS_crab} is a system of four Imaging Cherenkov Telescopes, located in the Khomas Highland of Namibia at an altitude of 1800\,m.
The \hessTel\ telescope array's location in the southern hemisphere is ideal for observing the Magellanic Clouds.
Furthermore, \hessTel\ is sensitive to gamma rays at energies above 100 \,GeV up to several 10\,TeV. The arrival direction of individual gamma rays can be reconstructed with an angular resolution of higher than $0.1\degr$, and their energy is estimated with an relative uncertainty of 15\%.

The region containing \nb\ was observed with \hessTel, the data presented having been recorded from 2004 to 2009, with a dead-time corrected exposure of 46
hours after data quality selection. The observations were carried out at large zenith angles (with a mean of $47^\circ$) leading to an elevated energy threshold of about 600\,GeV. The data were analysed using \textit{Model analysis} with standard cuts \citep{Mathieu}, where the camera images are compared with simulations using a log-likelihood minimisation. The remaining background was estimated from both rings around each sky position to generate the gamma-ray image and spatial analysis \citep[\textit{ring background}, ][]{BGmodels} and test regions with similar offsets from the camera centre for the spectral analysis \citep[\textit{reflected background}, ][]{BGmodels}.

Figure~\ref{Fig1}a shows a gamma-ray excess image of \nb. The image was smoothed with the \hessTel\ point-spread function, where 68\% of the events were contained
in a circle with a radius of $0.06\degr$. The significance was calculated from the counts in a circular integration region with a radius of $0.06\degr$ around each sky bin. Fitting a point-like source folded with the instrument's point spread function results in a best-fit position of the source of
RA = $5^{\mathrm{h}} 37^{\mathrm{m}} 44^{\mathrm{s}}$, Dec =
$- 69^{\circ} 9\arcmin 57\arcsec$, equinox J2000, with a statistical uncertainty of $\pm 11\arcsec$ in each direction; the source is hence
labelled {\hess}. With the H.E.S.S. standard pointing correction, point-like sources can be localised with a systematic uncertainty of $20\arcsec$ per axis \citepalias{GCposition}.
Figure~\ref{Fig1}b shows an X-ray image of the
supernova remnant and its central PWN with overlaid confidence contours of the gamma-ray source position. 
The best-fit position is consistent with the pulsar position; the slight offset from the pulsar along the tail of the PWN is not significant ($1.3\,\sigma$ of the combined statistical and systematic error).

In a circular region with a radius of $0.1\degr$ around the pulsar position, \Non\ gamma-ray candidate events were found. A total number of \Noff\ events were found in a background region with an area that is larger by a factor of \BGalpha. The corresponding gamma-ray excess is \Nex\ events with a statistical
significance of \Nsig\ \citep[calculated using formula (17) of][]{LiMa}. The photon spectrum between 600\,GeV and 12\,TeV of the gamma-ray excess can be
described by a pure power-law, $\mathrm{d}N/\mathrm{d}E = \Phi_0
\left(E/1\,\mathrm{TeV} \right)^{-\Gamma}$, with a normalisation at
1\,TeV of $\Phi_0 = $ \PhiZero\ and a photon index of $\Gamma = $ \SpecInd, which corresponds to an energy flux between 1\,TeV and 10\,TeV
of \EFlux\ or \CrabUnits\ of the energy flux of the Crab Nebula \citepalias{HESS_crab}. The gamma-ray luminosity between 1\,TeV and 10\,TeV for a distance of $48\,d_{48}$\,kpc is \Luminosity. This result was derived from data
accumulated over 4 years during the summer rainy seasons with varying
atmospheric conditions.  From the analysis of different subsets of the
data, a systematic uncertainty of 30\% in the flux and 0.3
in the photon index is estimated\footnote{These uncertainties are
  specific to this data set and should not be used for any other
  H.E.S.S. result.}.  

\section{Discussion}

The observed gamma-ray luminosity between 1\,TeV and 10\,TeV of the PWN corresponds to \efficiencyDist\ of the spin-down power of the pulsar, a value typical of young PWNe \citep[for a review see e.g.][]{Emma}.  Pulsar wind nebulae are very efficient in producing TeV gamma rays via the inverse Compton (IC) up-scattering of lower-energy photons by relativistic electrons\footnote{The term electrons refers to electrons and positrons.}. This makes them not only the most abundant source type among the TeV emitters in the Milky Way, but also with \hess\ the first extragalactic TeV source that is unrelated to a galaxy or an active galactic nucleus. \nb\ can be detected at such a large distance not only due to the high spin-down power of the pulsar but the strong infrared photon-fields from nearby sources serving as additional targets for the IC scattering.

\begin{figure}
   \centering
  \includegraphics[width=0.5\textwidth]{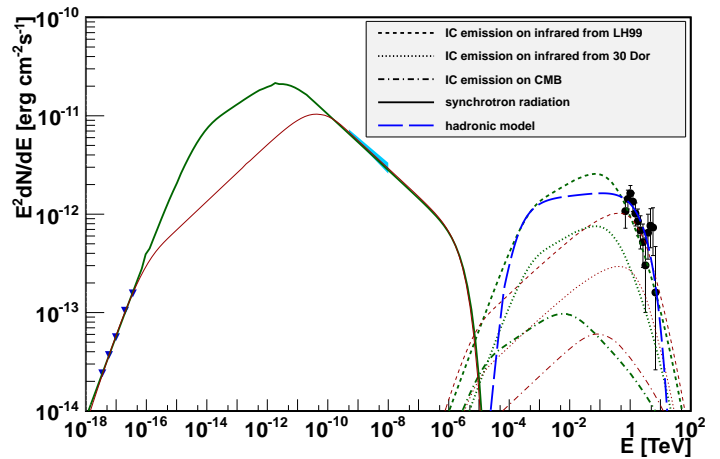}
      \caption{Spectral energy distribution of \nb. The data shown are
  radio emission \citep[][blue triangles]{Lazendic}, 
  non-thermal X-rays \citep[][blue band]{Chen},
   and TeV gamma rays (H.E.S.S., this
  work, black circles). The green, thick lines represent the model with a cooling break depending on the magnetic field and the age of the remnant. The red, thin lines denote the conservative model that minimises the energy content of electrons. The blue, long-dashed line represents a hadronic scenario.
              }
         \label{Fig2}
   \end{figure}

Figure~\ref{Fig2} shows the spectral energy distribution (SED) of \nb. It exhibits synchrotron emission that has been detected at radio wavelengths \citep[see][and references therein]{Lazendic}  and in X-rays \citep{Chen,XMM}, as well as TeV gamma-ray emission from IC scattering. In addition to the cosmic microwave
background (CMB), far infrared photons from the OB association LH\,99 and the nearby star-forming region 30 Doradus are important targets for IC scattering. Using observations from \textit{Spitzer} \citep{Spitzer}, the infrared photon fields are modelled as black-body radiation with a temperature of 80\,K and an energy density of $8.9\,\mathrm{eV\,cm}^{-3}$ for LH\,99, and a temperature of 88\,K and an energy density of $2.7\,\mathrm{eV\,cm}^{-3}$ for 30 Doradus. These are only upper limits to the infrared fields, as the (unprojected) distances of \nb\ to these objects are unknown.

There is no evidence of extended emission beyond the angular resolution
of H.E.S.S., thus the synchrotron and IC emission regions cannot be separated spatially.
Hence, a simple one-zone model is assumed, where only a single electron population is responsible for both the synchrotron and IC emission. An electron spectrum following a broken power-law with two breaks and covering the energy band from $1$\,eV to $10^{15}$\,eV is adopted. The low energy break is an intrinsic break of the injection spectrum proposed by \citet{VenterDeJager}; the high energy break arises from the cooling of the particles. It is assumed that the cooling break appears at the energy where the synchrotron loss time (which depends on the magnetic field) is equal to the age of the remnant (which is assumed to be 5000 years). A low-energy spectral index of $1.33$ is used to reproduce the radio spectrum, and a spectral index of $3.60$ above the cooling break is consistent with both the X-ray and TeV data. Assuming that the cooling steepens the spectral index by one \citep{Kardashev1962}, the uncooled high-energy spectral index is $2.60$.
This model, which is represented by the green lines in Figure~\ref{Fig2}, 
requires an intrinsic break at $54$\,GeV and a magnetic field of $41\,\mu$G, the corresponding cooling break being at $1.1$\,TeV. 
Assuming a distance of 48\,kpc, the total energy $W_\mathrm{tot}$ stored in electrons is $4 \times 10^{49}\,\mathrm{erg}$. In a second, more conservative model, any assumptions about the age and cooling of the particles are abandoned to minimise the total energy stored in electrons. The intrinsic break is set to 7\,GeV, the lowest energy still being compatible with the radio data, and the cooling break is set to 4.22\,TeV, the highest energy being compatible with the TeV data points. These prerequisites require an uncooled high-energy spectral index of $2.57$ and a low-energy spectral index of $0.93$ to reproduce the radio data. This model is represented by the red lines in Fig.~\ref{Fig2}; it requires a magnetic field of $47\,\mu$G and the total energy is only 50\% of the energy in the first model. The uncertainty in the distance measurement adds an error of 15\% to the estimate of $W_\mathrm{tot}$. Major uncertainties in the estimation of $W_\mathrm{tot}$ lie in the uncertainties in the gamma-ray spectrum.
Varying the TeV flux by 30\% while fixing the radio and X-ray points changes the total energy by 25\%.
The total energy content $W_\mathrm{tot}$ cannot be
determined from the radio and X-ray data alone, since the strength of
the synchrotron emission is governed by the \textit{a priori} unknown magnetic
field. Using the observed TeV spectrum presented here, it is possible to derive the magnetic field in the PWN and thus the energy content in electrons in the PWN.
In a hadronic scenario, gamma-ray emission is produced in the decay of $\pi^{0}$ mesons produced by inelastic interactions of accelerated protons with ambient material. The observed TeV emission can be described by the emission of a proton population following a power-law with an index 2 and an exponential cut-off at 23\,TeV (blue, long-dashed line in Fig.~\ref{Fig2}).  The total energy in this proton population is $W_{\mathrm{had}} = 1.8 \times 10^{52} (n/\mathrm{cm^{-3}})^{-1} d^2_{48} \mathrm{erg}$. 
An ambient density of at least $100\,\mathrm{cm}^{-3}$ would be necessary to produce this emission by a single supernova. That the SNR is expected to expand into the low-density interior of a superbubble makes this scenario unlikely.

As proposed by \citet{deJager2008}, $W_\mathrm{tot}$ can be used to estimate the birth period of the pulsar. \psr\ is relatively young, its characteristic age being shorter than the cooling time of most of the electrons. Particles have survived since the earliest epoch when the pulsar's spin was close to its birth period. Therefore, $W_\mathrm{tot}$ can  be related to the pulsar's birth and current period by the
calorimetric expression
\begin{eqnarray}
W_\mathrm{tot} 	& = & \epsilon \eta \left(E_\mathrm{rot,0} - E_\mathrm{rot}\right) \\
				& = & \epsilon \eta  \frac{1}{2} I \left(\left(\frac{2\pi}{P_0}\right)^2 - \left(\frac{2\pi}{P}\right)^2 \right) \\
    			& = & 2 \times 10^{50} \epsilon \eta \frac{I}{10^{45}\,\mathrm{g\,cm}^{2}} \left(\left(\frac{10\,\mathrm{ms}}{P_0}\right)^2 - \left(\frac{10\,\mathrm{ms}}{P}\right)^2 \right)\,\mathrm{erg},
\label{rel}
\end{eqnarray}
where $P_0$ is the birth period of the pulsar, $P$ is the
current period of $16.1 \,\mathrm{ms}$ \citep{Marshall1998}, $\eta$ denotes the conversion efficiency of spin-down power into accelerated
electrons, and $\epsilon$ is the relative average
energy-loss rate, which takes into account the energy already radiated
by the particles during earlier epochs and adiabatic losses of the energy. 
For the pulsar's moment of inertia, the canonical value of $I =
10^{45}\,\mathrm{g\,cm}^{2}$ is adopted \citep{LattimerPrakash2001}; the moment of inertia could be higher owing to variations in the pulsar mass between 1.4 solar masses ($M_\sun$) and 2.5 $M_\sun$ as proposed by \citet{Belczynski2008}.
Models indicate that $\eta$ can be as low as 0.3 \citep[for MSH 15-52]{Schoeck} or as high as 0.7 \citep[for G21.5$-$0.9]{G215}. Magnetohydrodynamical simulations \citep{deJager_ICRC} show that the adiabatic energy loss is around $\epsilon \sim 0.5$. Further radiation losses during earlier epochs of the nebula reduce $\epsilon$ to less than 0.5.  
Choosing 0.7 and 0.5 for $\eta$ and $\epsilon$, respectively, shows that the birth period must have been shorter than 10\,ms.
Using the conservative model, the birth period increases by 23\%. 
This result is consistent with earlier estimations of the pulsar's birth period, \citet{Marshall1998} estimated the birth period by comparing the pulsar's characteristic age with the age of the SNR for different braking indices, for SNR ages of more than 4000 years the birth period is shorter than 10\,ms for a large range of braking indices. From the extrapolation of glitch
data, \citet{Marshall2004} derive a pulsar period of 11\,ms 5000 years ago.
In the present paper, for the first
time the birth
period of a pulsar is obtained directly using a calorimetric technique, which depends on neither the glitch history nor the braking index and is --- for the conservative model --- completely independent of the age of the remnant.

This result confirms that \psr\ has with a birth period of shorter than about 10\,ms, the shortest birth period ever derived for a pulsar.
These short rotation periods are only known for millisecond pulsars that have been spun up after their birth by a companion star. 
Simulations show that pulsar birth periods can be related to some parameters of the progenitor stars. \citet{Heger2005} show that more massive progenitor stars produce
heavier and more rapidly rotating pulsars: 
stars of 15, 20 and 35 $M_\sun$ are required to produce pulsars with 11\,ms, 7\,ms and 3\,ms, respectively.
This is consistent with an earlier estimate of the \psr\ progenitor mass of 20-35 $M_\sun$ based on the comparison of the observed metal abundances in the
supernova ejecta of SNR \nb\ with supernova models \citep{Chen}. 
Such massive stars are close to the threshold for the formation of black holes in the supernova explosion. \citet{1999ApJ...522..413F} show, for instance, that 25 $M_\sun$ is roughly the limit for black hole formation.
On the other hand, \citet{Ott2006} show that the pulsar's birth period is rather unrelated to the progenitor's mass but roughly linearly dependent on the initial central iron-core spin. Birth periods of shorter than 10\,ms require initial iron-core periods of shorter than about 8\,s. A very massive and/or rapidly spinning progenitor star therefore appears to be required to produce a neutron star with a birth period as short as 10\,ms.

Alternatively, the pulsar could be part of a binary system that has been spun up by its companion star, a scenario typically assumed for millisecond pulsars. A very massive and rapidly rotating star at a distance of about $0.8\arcmin$ from \psr\ has been identified \citep{Dufton2011}. It was proposed that both objects were part of a binary system where mass was transferred  from the pulsar's progenitor to its companion and that both stars experienced radial velocity kicks in the supernova explosion. 
Nonetheless, in this scenario, the general picture of a very massive star producing a rapidly spinning neutron star remains unchanged.

\section{Conclusions}

Our principal conclusions are as follows:
   \begin{enumerate}
      \item Gamma-ray emission from the PWN in \nb\ was discovered with \hessTel\ observations. The energy flux between 1\,TeV and 10\,TeV is \EFlux. Located in the LMC at a distance of 48\,kpc, this is the most distant PWN ever detected in gamma rays and is the first individual stellar extragalactic TeV gamma-ray source. The PWN is powered by the most energetic pulsar known: \psr.
	 \item The TeV photon spectrum, in connection with radio and X-ray measurements, can be described with a one-zone leptonic model. From this model, the total energy stored in electrons in the nebula can be estimated to be $4\times10^{49}\mathrm{erg}$.
      \item For the pulsar to provide this energy from its rotational energy loss, the pulsar's birth period must have been shorter than about 10\,ms. This is the shortest birth period ever inferred for a pulsar. In an alternative scenario, the pulsar might have been spun up by a companion star in a binary system.
      \item Assuming a direct connection between the pulsar's birth period and the mass of the progenitor star, the progenitor must have had a mass of at least 15 $M_\sun$. This is close to the limit for black hole formation. The pulsar \psr\ is therefore at the upper mass limit for neutron star production.
   \end{enumerate}

\begin{acknowledgements}
The support of the Namibian authorities and of the University of Namibia
in facilitating the construction and operation of H.E.S.S. is gratefully
acknowledged, as is the support by the German Ministry for Education and
Research (BMBF), the Max Planck Society, the German Research Foundation (DFG), 
the French Ministry for Research,
the CNRS-IN2P3 and the Astroparticle Interdisciplinary Programme of the
CNRS, the U.K. Science and Technology Facilities Council (STFC),
the IPNP of the Charles University, the Czech Science Foundation, the Polish 
Ministry of Science and  Higher Education, the South African Department of
Science and Technology and National Research Foundation, and by the
University of Namibia. We appreciate the excellent work of the technical
support staff in Berlin, Durham, Hamburg, Heidelberg, Palaiseau, Paris,
Saclay, and in Namibia in the construction and operation of the
equipment.
\end{acknowledgements}

\bibliographystyle{aa} 
\bibliography{N157B}

\end{document}

%% file: includes.tex
\def\apjl{ApJ}%
\def\apjs{ApJS}%
\def\ao{Appl.~Opt.}%
\def\aap{A\&A}%
\def\degr{\hbox{$^\circ$}}
\def\arcmin{\hbox{$^\prime$}}
\def\arcsec{\hbox{$^{\prime\prime}$}}
\def\utw{\smash{\rlap{\lower5pt\hbox{$\sim$}}}}
\def\udtw{\smash{\rlap{\lower6pt\hbox{$\approx$}}}}

\def\sun{{\lower-2pt\hbox{$_\odot$}}}

\def\degr{\hbox{$^\circ$}}
\def\arcmin{\hbox{$^\prime$}}
\def\arcsec{\hbox{$^{\prime\prime}$}}

%% file: authorsAandA.tex
\author{H.E.S.S. Collaboration
\and A.~Abramowski \inst{1}
\and F.~Acero \inst{2}
\and F.~Aharonian \inst{3,4,5}
\and A.G.~Akhperjanian \inst{6,5}
\and G.~Anton \inst{7}
\and S.~Balenderan \inst{8}
\and A.~Balzer \inst{7}
\and A.~Barnacka \inst{9,10}
\and Y.~Becherini \inst{11,12}
% ...
}

\institute{
Universit\"at Hamburg, Institut f\"ur Experimentalphysik, Luruper Chaussee 149, D 22761 Hamburg, Germany \and
Laboratoire Univers et Particules de Montpellier, Universit\'e Montpellier 2, CNRS/IN2P3,  CC 72, Place Eug\`ene Bataillon, F-34095 Montpellier Cedex 5, France \and
Max-Planck-Institut f\"ur Kernphysik, P.O. Box 103980, D 69029 Heidelberg, Germany \and
Dublin Institute for Advanced Studies, 31 Fitzwilliam Place, Dublin 2, Ireland \and
National Academy of Sciences of the Republic of Armenia, Yerevan  \and
Yerevan Physics Institute, 2 Alikhanian Brothers St., 375036 Yerevan, Armenia \and
Universit\"at Erlangen-N\"urnberg, Physikalisches Institut, Erwin-Rommel-Str. 1, D 91058 Erlangen, Germany \and
University of Durham, Department of Physics, South Road, Durham DH1 3LE, U.K. \and
Nicolaus Copernicus Astronomical Center, ul. Bartycka 18, 00-716 Warsaw, Poland \and
CEA Saclay, DSM/IRFU, F-91191 Gif-Sur-Yvette Cedex, France \and
APC, AstroParticule et Cosmologie, Universit\'{e} Paris Diderot, CNRS/IN2P3, CEA/Irfu, Observatoire de Paris, Sorbonne Paris Cit\'{e}, 10, rue Alice Domon et L\'{e}onie Duquet, 75205 Paris Cedex 13, France,  \and
Laboratoire Leprince-Ringuet, Ecole Polytechnique, CNRS/IN2P3, F-91128 Palaiseau, France
% ...
}